\documentclass{ws-ijbc}

\usepackage[normalem]{ulem}

\usepackage{algorithmicx}

\usepackage{algpseudocode}
\usepackage{enumitem}
\usepackage{multicol}
\usepackage{ws-rotating}     
\usepackage{graphicx}
\usepackage{epstopdf}
\usepackage{tabularx}
\usepackage[titlenumbered,ruled,algo2e]{algorithm2e}

\usepackage{listings}

\usepackage{lmodern}
\usepackage[usenames,dvipsnames]{xcolor}

\usepackage{algorithm,algorithmicx}

\usepackage{algpseudocode}

\makeatletter
\newcommand\fs@nobottomruled{\def\@fs@cfont{\bfseries}\let\@fs@capt\floatc@ruled
  \def\@fs@pre{}%
  \def\@fs@post{}
  \def\@fs@mid{}%
  \let\@fs@iftopcapt\iftrue}
\makeatother

\lstnewenvironment{program}[2]{
    \renewcommand\lstlistingname{Program}
    \lstset{ ... }
} {}

\floatstyle{nobottomruled}
\restylefloat{algorithm}
\usepackage{xcolor}
\usepackage[toc,page]{appendix}
\usepackage{mathtools}
\usepackage{amsmath}%
\usepackage{amsfonts}%
\usepackage{amssymb}%
\usepackage{graphicx}
\usepackage{color}
\usepackage{cases}
\def\HiLi{\leavevmode\rlap{\hbox to \hsize{\color{yellow!50}\leaders\hrule height .8\baselineskip depth .5ex\hfill}}}

\newlist{myenumi}{description}{10}
\setlist[myenumi]{labelindent=\parindent, leftmargin=*, label=(\roman*), align=left}
\setlist[myenumi]{leftmargin=0pt}

\renewcommand{\theequation}{\arabic{section}.\arabic{equation}}

\def\({\left(}
\def\){\right)}

\newcommand{\Hilight}{\makebox[0pt][l]{\color{green}\rule[-4pt]{0.81\linewidth}{13pt}}}
\newcommand{\Hiligh}{\makebox[0pt][l]{\color{green}\rule[-4pt]{0.55\linewidth}{13pt}}}
\newcommand{\Hilig}{\makebox[0pt][l]{\color{green}\rule[-4pt]{0.72\linewidth}{13pt}}}
\newcommand{\Hili}{\makebox[0pt][l]{\color{yellow}\rule[-4pt]{0.8\linewidth}{13pt}}}

\definecolor{darkgreen}{rgb}{0.1, 0.5, 0.1}
\begin{document}
\title{Memory principle of the Matlab code for Lyapunov Exponents of fractional order}

\author{MARIUS-F. DANCA}

\address{STAR-UBB Institute, Bebes-Bolyai University,\\
Cluj-Napoca, Romania\\
and\\
Romanian Institute of Science and Technology, \\
Cluj-Napoca, Romania\\
m.f.danca@gmail.com}

\author{MICHAL FE\v{C}KAN}
\address{Department of Mathematical Analysis and Numerical Mathematics, \\Faculty of
Mathematics, Physics and Informatics, Comenius University in Bratislava, \\Bratislava, Slovakia;\\
and\\
Mathematical Institute of Slovak Academy of Sciences, \\Bratislava, Slovakia\\
michal.feckan@gmail.com}

\maketitle

\begin{history}
\received{(to be inserted by publisher)}
\end{history}

\begin{abstract}
The paper presents two representative classes of Impulsive Fractional Differential Equations defined with generalized Caputo's derivative, with fixed lower limit and changing lower limit, respectively. Memory principle is studied and numerical examples are considered. The problem of the memory principle of the Matlab code for Lyapunov exponents of fractional order systems \cite{dan1} is analyzed.
 \end{abstract}

\begin{multicols}{2}

\section{Introduction}Since the first conceptualization of fractional calculus initiated in correspondences in 1695 between Leibniz and L’Hopital and after the first definition of a fractional difference operator proposed in 1974 \cite{dia}, non-integer order differentiation and integration became an important field of research, in mathematics but also in many other parts of science, such as physics, biology, engineering and so on (see e.g. \cite{mati,kai2,apl1,apl2,doi} or monographies \cite{lak},\cite{kil},\cite{agar}) .

On the other hand, over the last years, the theory of impulsive differential equations suffered a
rapid development playing an important role in applied
mathematical models in physics, chemical technology, population dynamics, economy, biotechnology (see \cite{agar},\cite{ben},\cite{feck}).

Due to the intriguing non-locality property of fractional integrals and derivatives, solutions to FDEs require the entire time history having a consequence or memory meaning. The fractional order of a differential operator represents an important characteristic describing the quantity of the memory, the memory effect.
The existence of the output at the present moment of time depends on the history on a finite or infinite time interval. Therefore, compared with DEs where the solution is locally approximated, FDEs are extensively employed to depict the long-interaction phenomena (see \cite{tara} and references therein).

For performing numerical computation, the simplest approach is to disregard the tail of the memory and to integrate only over a fixed period of recent history. Therefore, one can be talked about short (fixed) principle \cite{doi} and, long (entire) memory principle (see e.g. \cite{wen,wen2} for effectiveness of the short memory principle, compared with long memory principle).

Regarding memory criteria in FDEs, in \cite{x} (see also \cite{xy}) the following paradigm is enounced: ``No memory. No fractional derivatives and integrals''.
The long time memory required by most  standard fractional derivatives, need large storage space in numerical simulations and cause poor efficiency. Therefore, short memory principles have been proposed to bring some fast numerical calculations.

On the other side, as very well known, the spectrum of Lyapunov Exponents (LE)s is a mathematical concept used to study dynamical systems, and to measure the rate of separation between two nearby trajectories in a system.

On one side, there are limitations and criticisms of using LEs due to their sensitivity to initial conditions, the difficulty to calculating accurately them for high-dimensional systems, or due to the fact that they also do not take into account the noise in a system or external factors. On the other side, because they continue to represent a numerical tool to determine a potential dynamic into a system several codes, especially for matlab, have been written (see e.g. \cite{wolf}. Generally the codes to determine LEs use the governing equations, or estimate them from time series.
While, because of the locally property of IO derivatives used to model DEs, the codes for LEs of IO are relatively easy to write, the situation changes significantly in the case of determining numerically the LEs of FO systems.

For LEs of autonomous systems of commensurate and incommensurate order, a set of two Matlab codes published in \cite{dan1}, \cite{dan2} have been written \cite{cod}. The codes adapts the integer order algorithm Benettin-Wolf \cite{bene1},\cite{bene2},\cite{wolf} to FDEs. For this purpose, the algorithm is transformed to integrate the underlying system of FDEs of commensurate and incommensurate order, respectively, together with the extended system, with a fixed step-size numerical scheme (a fast Adams-Bashforth-Moulton (ABM) scheme for FDEs) as explained in \cite{dan1} and \cite{dan2}.
It is shown that the codes enjoy a long (entire) memory principle.

The main purpose of the paper is two folds. Firstly, following the reasonings used to determine the integral of a general impulsive system of DEs, presented in Section 2, two classes of impulsive FDEs modeled with a generalized Caputo's derivative are considered in Section 3, with fixed lower limit and changing lower limit. Also, the memory principle is analyzed and underlined by examples. Secondly, the memory principle of the Matlab code for LEs of fractional order systems \cite{cod} is analyzed in Section 4.

The numerical method utilized in the considered examples of FO is the  Adams-Bashforth-Moulton (ABM) scheme for FDEs with fixed step size $h=0.02$ \cite{kai2}.

\section{Impulsive Differential Equations of Integer Order}\label{sect1}

Consider the Cauchy problem for the following impulsive differential equations

\begin{equation}\label{e10}
	\begin{aligned}
		&x'(t) = f(t,x(t)),\quad t \in J':=J\setminus \{t_1,t_2,\ldots t_m\}, \\
        &J=[0,T],\\
		&x(t_k^+) = x(t_k^-)+y_k,\quad k=1,2,\ldots,m,\\
		&x(t_0) = x_0,
	\end{aligned}
\end{equation}
$f:\mathbb{R}\times \mathbb{R}^n\rightarrow \mathbb{R}^n$, $x_0\in \mathbb{R}^n$, $t_0<t_1<\ldots<t_{m+1}=T$, $x(t^+)=\lim_{\epsilon \rightarrow 0^+}(x(t_k+\epsilon)$ and $x(t^-)=\lim_{\epsilon \rightarrow 0^-}(x(t_k+\epsilon)$ represent the right and left limits of $x(t)$ at the impulse time $t=t_k$, $k=1,2,...,m$, $m\in \mathbb{Z}^+$ and $T\in \mathbb{R}^+$.
Impulses are considered at moments $t_1,...,t_m$.

\noindent The following mild assumption is considered

\noindent(\textbf{H}) The function $f$ is continuous

Let first solve the following differential equation by using some numerical scheme

\begin{equation}\label{e1}
x'(t)=f(t,x(t)),\, t\in[t_k,t_{k+1}),
\end{equation}
with impulse
\begin{equation}\label{e2}
x(t_k^+)=x(t_k^-)+y_k,
\end{equation}
for $k=0$ with $x(t_0)=x_0$ at $t_0$, for $t\in[t_0,t_1)$.

\noindent For $k=1$, we use the impulse \eqref{e2} to get
\[
x(t_1^+)=x(t_1^-)+y_1,
\]
 and we solve \eqref{e1} on $[t_1,t_2)$, and so on. On the other hand, \eqref{e1} gives
\begin{equation}\label{e3}
\begin{gathered}
x(t_1^-)=x(t_0)+\int_{t_0}^{t_1}f(s,x(s))ds,\\
x(t)=x(t_1^+)+\int_{t_1}^{t}f(s,x(s))ds,\, t\in[t_1,t_{2}).
\end{gathered}
\end{equation}
Using \eqref{e2} and \eqref{e3}, we arrive at
$$
\begin{gathered}
x(t)=x(t_1^+)+\int_{t_1}^{t}f(s,x(s))ds\\
=x(t_0)+\int_{t_0}^{t_1}f(s,x(s))ds+y_1+\int_{t_1}^{t}f(s,x(s))ds\\
=x(t_0)+y_1+\int_{t_0}^{t}f(s,x(s))ds.
\end{gathered}
$$
for $t\in[t_1,t_{2})$. So
$$
x(t)=x(t_0)+y_1+\int_{t_0}^{t}f(s,x(s))ds,\, t\in[t_1,t_{2}).
$$
By repeating this argument for any $k$ one get the integral over the entire time integration interval $J$
\begin{equation}\label{e4}
\begin{split}
&x(t)=x(t_0)+\sum_{i=1}^ky_i+\int_{t_0}^{t}f(s,x(s))ds,\, t\in[t_k,t_{k+1}),~~~\\
&k=1,2,\ldots,m.
\end{split}
\end{equation}

\section{Impulsive fractional differential equations}

In this section, two impulsive FDEs are considered, which will be called hereafter \emph{impulsive algorithms} (see \cite{ultima}).

\subsection{Impulsive differential equations with fixed lower limit}

Let the fractional variant of the impulsive problem \eqref{e10}
\begin{equation}\label{pp}
\begin{split}
&^{c}D_{t_0}^{q}x(t) = f(t,x(t)),\quad t \in J',\\
&x(t_k^+) = x(t_k^-)+y_k,\quad k=1,2,\ldots,m,\\
		&x(t_0) = x_0,
\end{split}
\end{equation}
where $^{c}D_{t_0}^{q}x(t)$ stands as the generalized Caputo fractional derivative with Fixed Lower Limit (FLL) at $t_0$, of fractional order $q\in(0,1)$.

By applying the reasoning used the Section \ref{sect1} one obtains the solution of the impulsive FLL algorithm \eqref{pp} (see also \cite{trei})
\begin{equation}\label{e8}
\begin{split}
&x(t)=x(t_0)+\sum_{i=1}^ky_i+\frac{1}{\Gamma(q)}\int_{t_0}^{t}(t-s)^{q-1}f(s,x(s))ds,~~\\
&t\in[t_k,t_{k+1}),\quad k=1,2,\ldots,m.
\end{split}
\end{equation}

\noindent For $q=1$, the formula \eqref{e8} coincides to the integer order formula \eqref{e4}.

\vspace{3mm}
\begin{remark}\itshape\label{cont}
Note that by the continuity of solution on $[t_k,t_{k+1}]$ (see e.g. \cite{kai}), in applications, the upper limit of integrals, $t$, can be taken on the right side of the interval, $t=t_{k+1}$.
\end{remark}
\vspace{3mm}

\begin{remark}\itshape\label{remi}
\begin{itemize}[topsep=0pt,itemindent=.2in]
\item[i)]When the lower limit is fixed at $t_0$, the underlying FDEs have no nonconstant periodic solutions \cite{tava}.
\item[ii)] If in the impulsive FLL algorithm the nodes $t_k$ are periodic, we have a dynamical system, so there are equilibria, periodic points etc. \cite{FWZ,FW};
\end{itemize}
\end{remark}
\vspace{3mm}
For $t\in[t_k,t_{k+1})$, in the non-impulsive case ($y_k=0, k=1,2,\ldots,m$), the formula \eqref{e8} becomes
\begin{equation}\label{ee1}
\begin{split}
&x(t)=x(t_0)+\frac{1}{\Gamma(q)}\int_{t_0}^{t}(t-s)^{q-1}f(s,x(s))ds,\\
&t\in[t_k,t_{k+1}),
\end{split}
\end{equation}
and, therefore, the following property holds
\begin{proposition}
For $t\in[t_k,t_{k+1})$, $k=1,2,\ldots,m$, the solution \eqref{e8} of the impulsive FLL algorithm \eqref{pp}, represents a translation in $\mathbb{R}^n$ of the non-impulsive solution \eqref{ee1} with the term $\sum_{i=1}^ky_i$.

\end{proposition}

Moreover, as will be seen in the following example (Example 1), the shape of the generated impulsive trajectories are actually not influenced by the impulses, but suffer a translation only and, therefore, the usefulness in practical applications of the impulsive FLL algorithm can be questioned.

 In this paper for simplicity, in examples, identic impulses $y_k=y$, $k=1,2,\ldots,m$ are considered, and the nodes $t_k$ are supposed equally spaced. The ABM scheme is used.

\begin{figure}[ptbh]
 \includegraphics[width=1.1\textwidth]{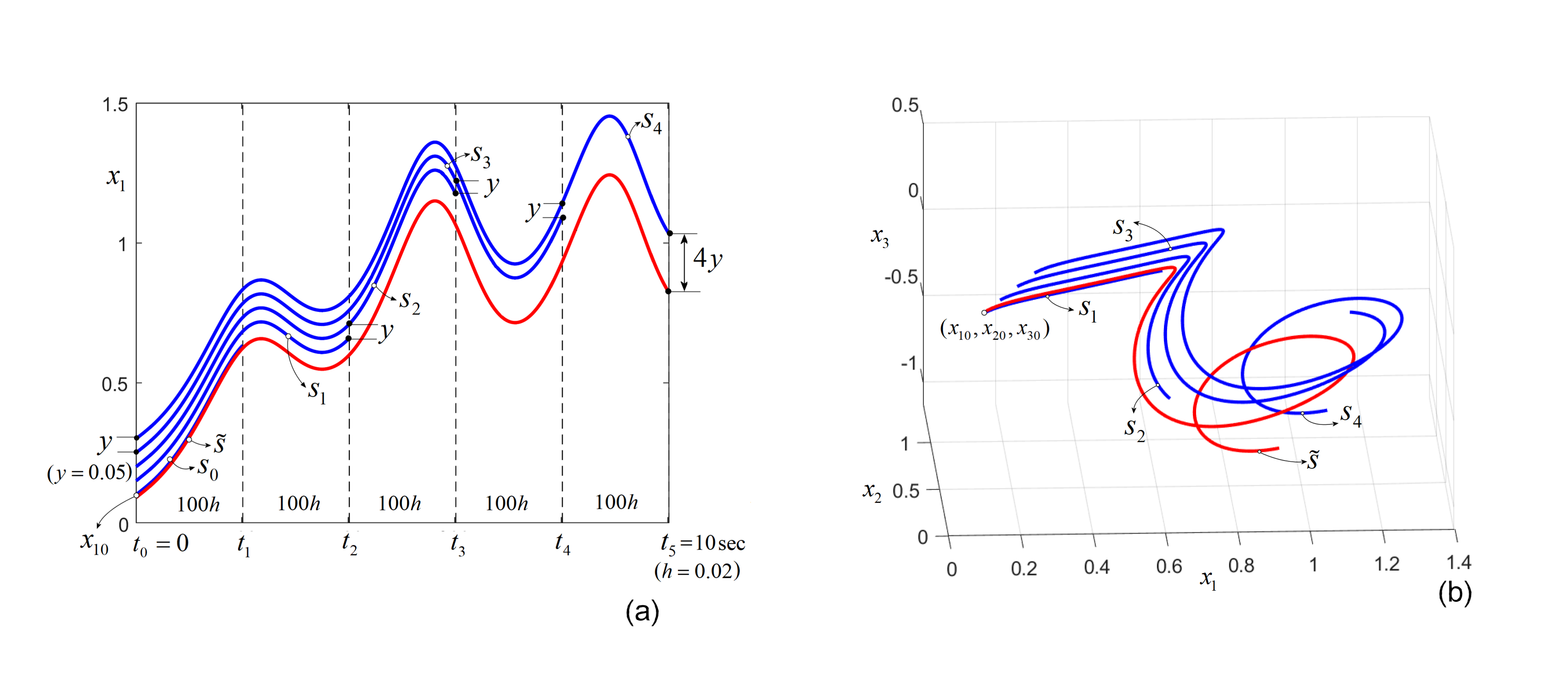}
\caption{Impulsive FLL algorithm applied to system \eqref{ecc} on four time intervals. $\tilde{S}$ is the non-impulse trajectory, $S_1,$,...,$S_4$ are impulse trajectories, while $S_0$ is non-impulsive trajectory; (a) Time series of variable $x_1$; (b) Phase plot.}
\label{fig1}       
\end{figure}

\begin{itemize}[itemindent=.77in]\label{ex1}
\item[\emph{Example 1}]

Consider the fractional variant of the system  modeling the interaction between dark matter and dark energy \cite{dan3}

\begin{equation}\label{ecc}
\begin{split}
^{c}D_0^{q}x_1&=x_2x_3-x_1,\\
^{c}D_0^{q}x_2&=(x_3-5)x_1-x_2,\\
^{c}D_0^{q}x_3&=1-x_1x_2,
\end{split}
\end{equation}

whose rich dynamics include a hidden chaotic attractor for $q=0.995$.

By using the impulsive FLL algorithm, with $(x_{10},x_{20},x_{30})=(0.1,0.1,0.1)$, and $T=[0,10sec]$ in the particular case of identical impulses, $y=0.05$, applied at four moments $t_k=100kh$, where $h=0.02sec$ is the step size of the ABM method, one obtain the trajectories given by the integrals \eqref{e8} denoted $S_k, k=1,2,3,4$ in Figs. \ref{fig1} (in Fig. \ref{fig1} (a) the time series of variable $x_1$ is presented (blue plot), while in Fig. \ref{fig1} (b), for clarity, phase portrait of the only first three impulsive trajectories are presented). As can be seen, the impulsive trajectories $S_k$ are translated in with $\sum_{i=1}^ky=ky=k0.05$, $k=1,2,3,4$ from the corresponding segments of the non-impulsive trajectory $\tilde{S}$ (red plot in Figs. \ref{fig1}). Obviously, if the impulses $y$ are zero, the solution is identical with $\tilde{S}$ ($S_0$, for $t\in [0,t_1]$).


\end{itemize}

\subsection{Impulsive fractional differential equation with changing lower limit }
To avoid the obstacle in using numerically the impulsive FLL algorithm (see Example 1), consider next impulsive FDE \eqref{e10} with the Changing Lower Limit (CLL) at $t_k$, $k=1,2,\ldots,m$
\begin{equation}\label{e101}
	\begin{split}
		&^{c}D_{t_k}^{q}x(t) = f(t,x(t)),\quad t \in J',\\
		&x(t_k^+)= x(t_k^-)+y_k,\quad k=1,2,\ldots,m,\\
		&x(t_k) = x_k,
	\end{split}
\end{equation}
Using the reasoning in \eqref{e3} one obtains the solution on $t\in[t_k,t_{k+1})$

\begin{equation}\label{e6}
\begin{split}
&x(t)=x(t_k^+)+\frac{1}{\Gamma(q)}\int_{t_k}^{t}(t-s)^{q-1}f(s,x(s))ds,
\end{split}
\end{equation}
where the value $x(t_k^+)$ is the value $x(t_k^-)$ modified by the impulse $y_k$
\begin{equation}\label{imp}
 x(t_k^+)=x(t_k^-)+y_k.
 \end{equation}

\noindent Therefore, by gathering relations \eqref{e6} one obtains finally the expression on $t\in[t_k,t_{k+1})$

\begin{equation}\label{ttt}
\begin{split}
&x(t)=x(t_k^+)+\frac{1}{\Gamma(q)}\int_{t_k}^{t}(t-s)^{q-1}f(s,x(s))ds\\
&=x(t_k^-)+y_k+\frac{1}{\Gamma(q)}\int_{t_k}^{t}(t-s)^{q-1}f(s,x(s))ds\\
&=x(t_{k-1}^-)+\frac{1}{\Gamma(q)}\int_{t_{k-1}}^{t_k}(t_k-s)^{q-1}f(s,x(s))ds\\
&+y_k+\frac{1}{\Gamma(q)}\int_{t_k}^{t}(t-s)^{q-1}f(s,x(s))ds\\
&=x(t_0)+\sum_{i=1}^ky_i+\frac{1}{\Gamma(q)}\sum_{i=1}^k\int_{t_{i-1}}^{t_i}(t_i-s)^{q-1}f(s,x(s))ds\\
&+\frac{1}{\Gamma(q)}\int_{t_k}^{t}(t-s)^{q-1}f(s,x(s))ds,
\end{split}
\end{equation}
which represents the solution of the impulsive CLL algorithm \eqref{e101} for $t\in[t_k,t_{k+1})$.

As for all impulsive schemes, the solution of the impulsive algorithms FLL and CLL are piece-wise continuous.

\noindent To note that in \eqref{ttt}
\begin{equation*}\label{e7}
\begin{split}
&\int_{t_0}^{t}(t-s)^{q-1}f(s,x(s))ds\ne \\
&\sum_{i=1}^k\int_{t_{i-1}}^{t_i}(t_i-s)^{q-1}f(s,x(s))ds+\\
&\int_{t_k}^{t}(t-s)^{q-1}f(s,x(s))ds,
\end{split}
\end{equation*}

\begin{figure}
  \includegraphics[width=.5\textwidth]{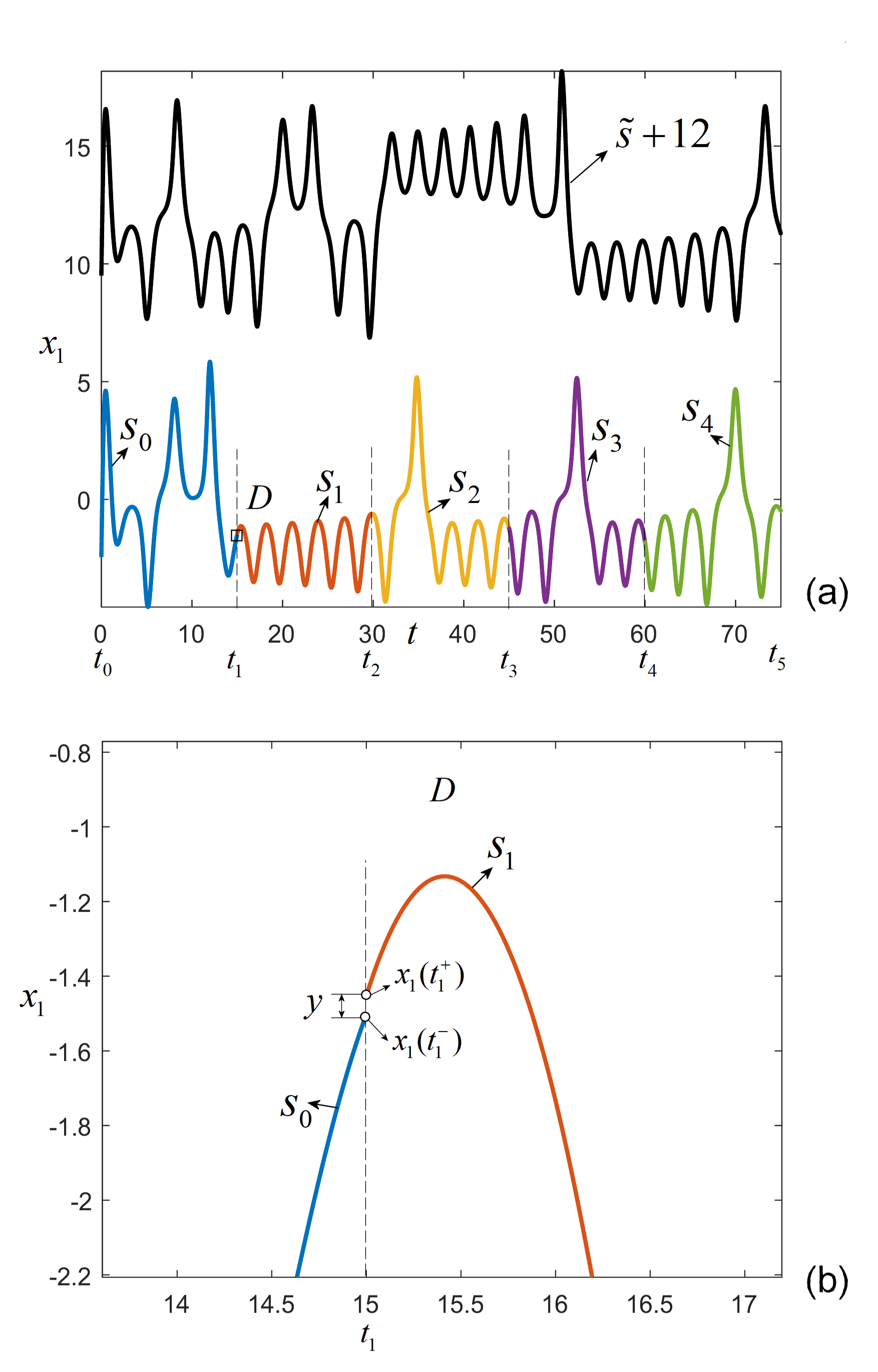}
\caption{Impulsive CLL algorithm applied to system \eqref{ecc} on five time intervals; (a) The five different obtained trajectories $S_i$, $i=1,2,3,4$ are colored ($S_0$ is non-impulsive trajectory); (b) Zoomed detail $D$ showing the impulse discontinuity.}
\label{fig2}       
\end{figure}

\begin{itemize}[itemindent=.77in]\label{ex2}
\item[\emph{Example 2}]
If one apply the impulsive algorithm CLL  \eqref{e101} to the system \eqref{ecc} with $q=0.995$ on four equidistant time moments $t_k=15k, k=1,2,3,4$, one obtains the coloured time series of the state $x_1$, denoted $S_0,S_1,\ldots, S_4$, on Fig. \ref{fig2} (a). For clarity, the non-impulsive trajectory $\tilde{S}$ (black plot) is shifted along the vertical axis with 12 along the $x_1$. The considered impulses along the three axes are $y_1=y_2=y_3=0.05$ and the inherent discontinuity of the solution is unveiled by the rectangular zoom  $D$ in Fig. \ref{fig2} (b), at the node $t_1=15$ (as in Example 1, $S_0$ represents the solution on $[t_0,t_1)$, where there are no impulses).
\end{itemize}

\begin{figure}
  \includegraphics[width=.8\textwidth]{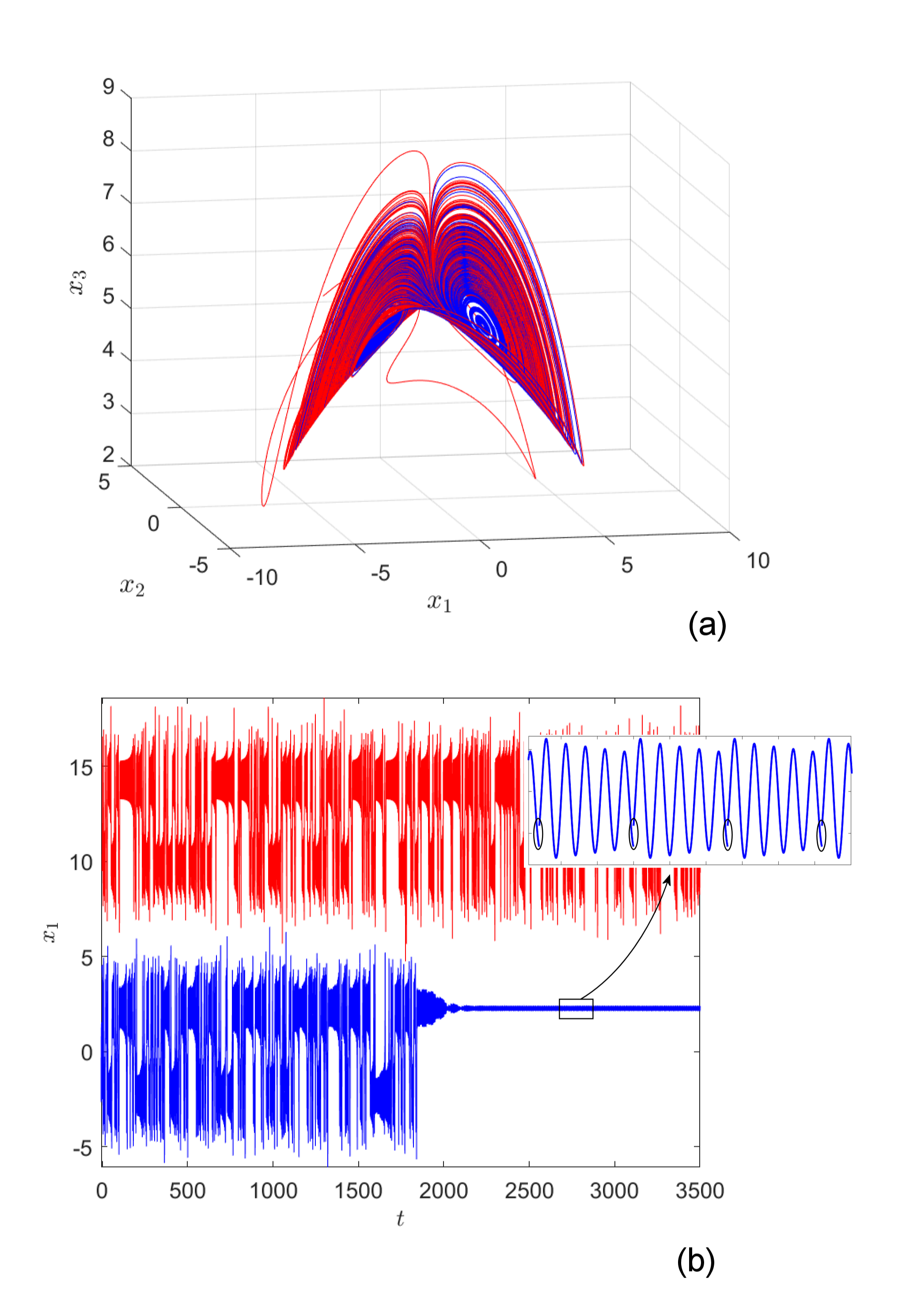}
\caption{Overplotted chaotic trajectories of system \eqref{ecc} obtained with the impulsive (blue plot) algorithm and non-impulsive (red plot) trajectory; (b) Time series revealing the fact that, after a relative long chaotic transient, the chaotic behavior can be transformed into a regular-like one. The zoomed detail shows, beside the non-periodicity, the discontinuity due to impulses. }
\label{fig3}       
\end{figure}

\begin{itemize}[itemindent=.77in]\label{ex3}
\item[\emph{Example 3}]
Note that the impulsive CLL algorithm \eqref{e101} can be used as a control-like scheme. Consider the same example of the system \eqref{ecc}, without impulses. For a sufficient long integration time, $[0,3500]$, a chaotic trajectory is generated (see red plot in Figs \ref{fig3} (a) and (b)). If one considers the impulsive system with impulse $y=0.05$ applied at every moment $t_k=13kh$, $k=1,2,\ldots$, after a sufficient long transient, at about $t=1700$, a controlled-like trajectory (see Remark \ref{remi}) is obtained (blue plot in the phase plot in Fig. \ref{fig3} (a) and time series of $x_1$ in Fig. \ref{fig3} (b)). As the zoomed detail reveals, as expected, the obtained trajectory is discontinued and is composed by a train of bursts.
\end{itemize}

As known, the nonlocality by time of fractional derivatives, implies a dynamic memory process. Mathematically, the property of nonlocality is expressed in the fact that fractional derivatives of non-integer
orders cannot be represented as a finite set of derivatives of integer orders \cite{x}.

\begin{figure}[ptbh]
 \includegraphics[width=.65\textwidth]{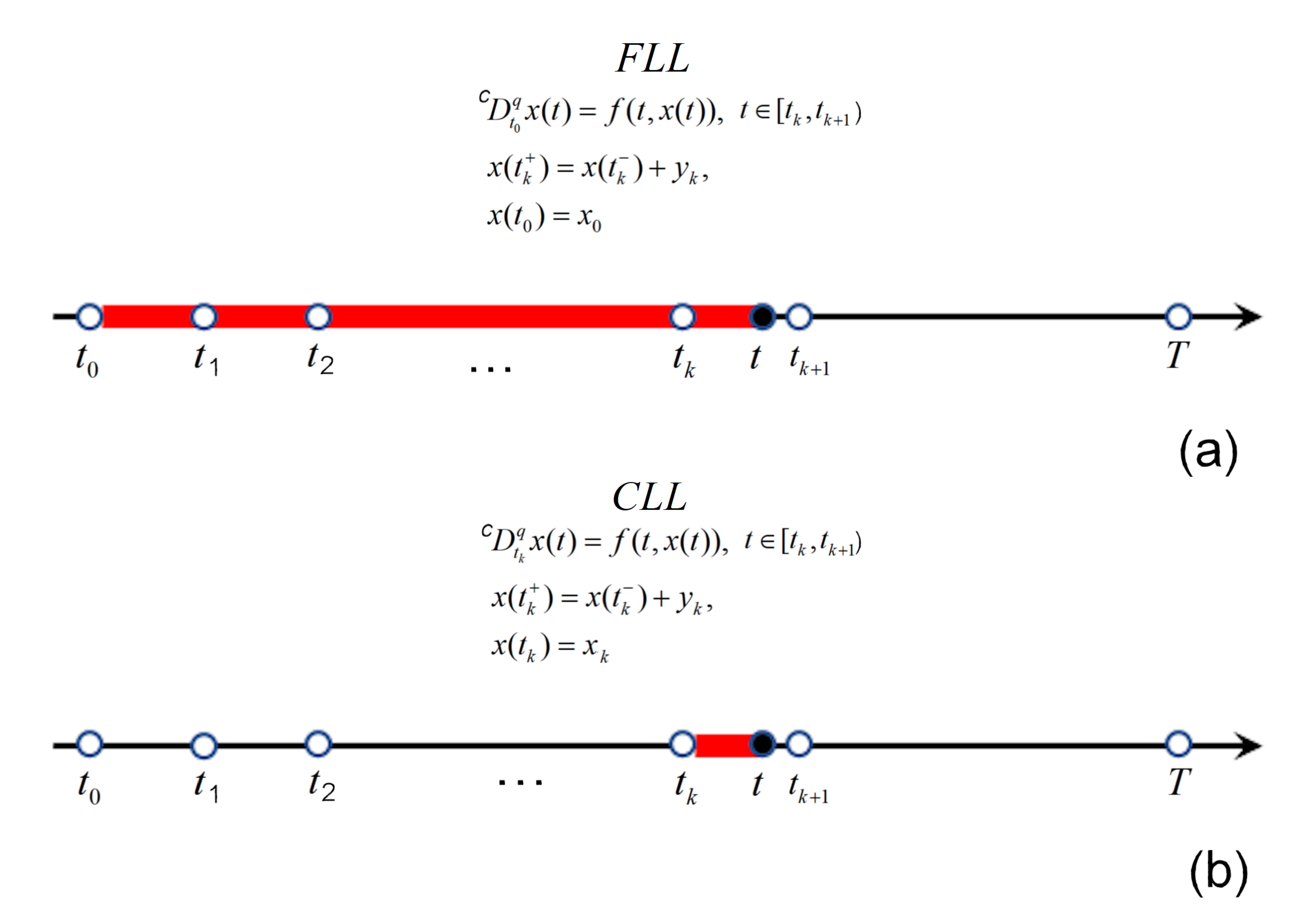}
\caption{Sketch of the memory principles (red plot); (a) FLL algorithm; (b) CLL algorithm.}
\label{fig4}       
\end{figure}

Formulae of the solutions of the impulsive FLL and CLL algorithms, \eqref{e4} and \eqref{ttt}, respectively, look similar but, however, they are different. Thus, both algorithms are \emph{full (entire) memory principle}, in the sense that the unknown value $x(t)$ depend on \emph{all previously determined values}, but the impulsive FLL algorithm considers the memory from the lower limit $t_0$, while in the case of the impulsive CLL algorithm, the memory is considered from the lower limit $t_k$ (see the sketch in Fig. \ref{fig4}). The time length of the integrals are $t-t_0$ and $t-t_k$, respectively, for $k=1,2,\ldots,m$.

The considered algorithms FLL and CLL enjoy the following property

\begin{proposition}
The impulsive FLL and CLL algorithms are full memory.
\end{proposition}

\begin{remark}\itshape
\begin{itemize}[topsep=0pt,itemindent=.3in]
\item[i)] Both FLL and CLL variants are not shorth (fixed) memory length, when the tail of the integration is disregarded so that the length of the memory remain constant (see e.g. \cite{doi}). Considering the CLL algorithm, even it looks like a short memory principle, finally, on the considered interval $[t_k,t_{k+1})$,  the memory segment $t-t_k$, considered from $t_k$, increases as $t$ approaches $t_{k+1}$.

    \item [ii)]Despite the fact that solving FDEs with full memory principle is computationally time consuming, needing large storage space, the computational errors are relatively small.
  \item [iii)]As shown by the considered Examples 1, 2 and 3, compared with the impulsive CLL algorithm, where the impulses have a substantial role in system's behavior, the impulses in the impulsive FLL algorithm, does not affect the dynamics of the underlying system, the influence of impulses being only to translate the non-impulsive trajectories.
\end{itemize}
\end{remark}

\section{Memory principle in Matlab code for LEs of fractional order systems}
In \cite{dan1} and \cite{dan2} are introduced the matlab codes to calculate the spectrum of LEs of fractional systems of commensurate and incommensurate order respectively, some of the few existing matlab code for LEs of FO. The codes are made by adapting the effective algorithm for computing the spectrum of LEs for IO presented in 1980 by Benettin and others \cite{bene1}, \cite{bene2}. Updated versions and related codes can be found at \cite{cod} (see also Appendix).

In this paper we answer to the open question if these codes present memory principle.

Consider the case of the code for fractional commensurate order $FO\text\_Lyapunov.m$ \cite{dan1}, for the incommensurate order \cite{dan2} the reasoning being similar.

The algorithm uses the Gram-Schmidt (G-S) process which, given some arbitrary vector basis $\{u_1,u_2,\ldots,u_n\}$ of $\mathbb{R}^n$, forms a new orthonormal basis $\{v_1,v_2,\ldots,v_n\}$ of $\mathbb{R}^n$ from it. It is designed to turn a basis into an ortho-normal basis without altering the subspace that it spans. Going one by one through the vectors, subtract off the part not orthogonal to the previous ones. Next, each vector are normalized in the resulting basis by dividing it by its norm (see the matlab code $gramschmidt.m$, in Algorithm \ref{ddd}, which simulates the G-S procedure \url{https://en.wikipedia.org/wiki/Gram-Schmidt_process}). The first (green) line 8 subtracts off the part not orthogonal vectors to the previous ones while the second line 10 normalizes the resulting vectors.

As presented in \cite{dan1}, in the $FO\text\_Lyapunov.m$ code, the G-S algorithm acts periodically on the equidistant nodes $t_k,=kh_{norm}$, $k=1,2,\ldots$, where $h_{norm}$ determines the normalization
moments (see the sketch in Fig.\ref{fig5}).

To start the program the following command is used
\lstset{numbers=none}
\begin{lstlisting}
[t,LE]=FO_Lyapunov(ne,ext_fcn,t_start,h_norm,t_end,x_start,h,q,out)
\end{lstlisting}
\vspace{3mm}
where $ne$ represents the number of the equations modelling the system, $ext\_fcn$, the function containing the equations and the variational equations, $t\_start$ and $t\_end$, determine the interval of integration (here realized with the ABM matlab code FDE12 \cite{fde12}), $h\_norm$, determines the moments when the G-S algorithm applies, $x\_start$, are the initial conditions, $h$, is the step of the numerical integrator, $q$, represent the commensurate fractional order, and $out$ represents the display moments (examples can be found on \cite{cod} or \cite{dan1}).

For example, in the case of the system \eqref{ecc}, the command
\begin{lstlisting}
[t,LE]=FO_Lyapunov(3,@dark,0,0.1,1000, [0.1;0.1; 0.1],0.02,0.995,500);
\end{lstlisting}
where, $h\_norm=0.1$, $h=0.02$, and the function $dark.m$ given in Appendix, produces the following result (see also Fig. \ref{figLE} where LEs are plotted)
\begin{lstlisting}
     50.00     0.2919    -0.0546    -2.5591
    100.00     0.2604    -0.0229    -2.4680
    150.00     0.2538    -0.0151    -2.4014
    200.00     0.2543    -0.0086    -2.3743
    ...
    900.00     0.2552    -0.0029    -2.5898
    950.00     0.2538    -0.0019    -2.5942
   1000.00     0.2561    -0.0025    -2.6003
\end{lstlisting}
at the end of integration, $t=1000$, LEs being $(0.2561,-0.0025, -2.6003)$. Regarding the potential relatively significant errors, which characterize not only this algorithm, but all numerical algorithms for LEs, see \cite{dan1}.

The two G-S phases of orthogonalization and normalization are underlined by green color, respectively, in the code in Appendix (compare with lines 8 and 10 in Gram-Schmidt code in Listing 1). The integrator integrates step by step (of length $h$), simultaneously the considered system and its variational equations (see \cite{dan1}), and every moments $kh_{norm}$ (multiple of $h$), for $k=1,2,\ldots$, the G-S procedure being applied.

On the other side, the yellow colored lines in the code (Appendix)
\vspace{3mm}
\lstset{numbers=none}
\begin{lstlisting}
[T,Y] = FDE12(q,ext_fcn,t,t+h_norm,x,h);
t=t+h_norm;
\end{lstlisting}
\vspace{3mm}
indicate that the numerical integration acts with the step size integration $h$ (argument in the FDE12 routine), until the interval $[t,t+h_{norm}]$ (second line) is covered by integration. In the sketch in Fig.\ref{fig5} $h_{norm}=2h$ (see \cite{dan1} for the choice of $h_{norm}$ as function of $h$). At the end of these intervals of length $h_{norm}$, the G-S procedure (green lines) is applied. The previously obtained integration value, $x$, is used as initial condition for the next integration (argument in the FDE12 routine) within $[t,t+h_{norm})$, after which is modified by the G-S procedure.

Therefore, the algorithm for LEs of FO is impulsive, applied in all intervals $[t_k,t_{k+1})=[kh_{norm},(k+1)h_{norm})$, $k=1,2,\ldots$, successively (Fig. \ref{fig5}).

\noindent The following property follows

\begin{proposition}
The matlab code for FO LEs is an impulsive CLL algorithm, with variable (entire) memory.
\end{proposition}

Fixed lower impulsive algorithms, like the FLL \eqref{pp}, cannot be applied in this case (see Example \ref{ex1}).

\section*{Conclusion}
In this paper there are presented two approaches of impulsive fractional differential equations modeled by generalized Caputo's fractional derivative, with fixed lower limit defined at $t_0$, and with changing lower limit at nodes $t_k$, respectively. Both procedures enjoy the full memory principle which even are more time consuming compared to fixed memory principles, take the advantage of a higher precision. While the fixed lower limit approach seems less useful in some practical applications, the changing lower limit approach proves to be more useful in practical applications, as the considered examples show. It is shown that the Matlab code for LEs of fractional-order systems \cite{dan1} (downloadable at \cite{cod}), is an impulsive changing lower limit algorithm with full memory principle. Further studies such as a comparative study regarding computational time of both methods represent a future task.
\vspace{5mm}

\textbf{Acknowledgement. }This paper is partially supported by the Slovak Grant Agency VEGA No. 1/0084/23 and No. 2/0062/24.
\end{multicols}

\newpage

\mbox{}
\vspace{.1cm}
\begin{multicols}{2}
\lstset{numbers=none
}
\section*{Appendices}

Matlab code for LEs of fractional order systems

\begin{lstlisting}[title=dd,captionpos=b,escapechar=|,tabsize=1]
function [t,LE]=FO_Lyapunov(ne,ext_fcn,t_start,h_norm,t_end,x_start,h,q,out);
%
% Program to compute LEs as function of time of
% autonomous systems of commensurate Fractional
% Order defined via Caputo's derivative;
%
% author: Marius-F. Danca, August 2022
% email: m.f.danca@gmail.ro
%
% References, details on the code and auxiliary
% files can be found at
%
%//https://www.mathworks.com/matlabcentral/profile/
% authors/3083566
%
% How to use it:
% [t,LE]=FO_Lyapunov(ne,ext_fcn,t_start,h_norm,
% t_end,x_start,h,q,out);
%
% Input:
% -ne - system dimension (eqs number);
% -ext_fcn - function containing the extended
% system of FO;
% -t_start, t_end - time span;
% -h - integration step;
% -h_norm - step for Gram-Schmidt renormalization.
% Optimal choice: multiple of h;
% -x_start - initial condition;
% -q - the fractional order;
% -out - priniting step of LEs values;
% out==0 - no print.
%
% Output:
% t - time values;
% LE -  Lyapunov exponents printed every 'out' step.
%
% Example of use for the RF system:
% [t,LE]=FO_Lyapunov(3,@LE_RF,
% 0,0.02,300,[0.1;0.1;0.1],0.02,0.98,1000);
%
% Cite as:
%
% Marius-F. Danca and N. Kuznetsov, Matlab code for
% Lyapunov exponents of fractional order systems,
% International Journal of Bifurcation and Chaos,
% 28(05)(2018), 1850067
%
% see also:
% Marius.-F. Danca, Matlab code for Lyapunov
% exponents of fractional-order systems, Part II:
% The non-commensurate case, International Journal
% of Bifurcation and Chaos, 31(12), 2150187, (2021)
%
hold on;
colors = 'grby';%to include any of the 8 color
% codes bcgkmrwy
% Memory allocation
x=zeros(ne*(ne+1),1);
x0=x;
c=zeros(ne,1);
gsc=c; zn=c;
n_it = round((t_end-t_start)/h_norm);
% Initial values
    x(1:ne)=x_start;
    i=1;
    while i<=ne
        x((ne+1)*i)=1.0;
        i=i+1;
    end
    t=t_start;
% Main loop
it=1;
while it<=n_it
   LE=zeros(ne,1);
    % Solutuion of extended ODE system of FO using FDE12 routine
   |\Hili| [T,Y] = FDE12(q,ext_fcn,t,t+h_norm,x,h);
   |\Hili| t=t+h_norm;
    Y=transpose(Y);
    x=Y(size(Y,1),:); %solution at t+h_norm
    i=1;
    while i<=ne
        j=1;
        while j<=ne;
            x0(ne*i+j)=x(ne*j+i);
            j=j+1;
        end;
        i=i+1;
    end;
%  construct new orthonormal basis by
%  gram-schmidt procedure
    zn(1)=0.0;
    j=1;
    while j<=ne
        zn(1)=zn(1)+x0(ne*j+1)^2;
        j=j+1;
    end;
    zn(1)=sqrt(zn(1));
    j=1;
    while j<=ne
        x0(ne*j+1)=x0(ne*j+1)/zn(1);
        j=j+1;
    end
    j=2;
    while j<=ne
        k=1;
        while k<=j-1
            gsc(k)=0.0;
            l=1;
            while l<=ne;
                gsc(k)=gsc(k)+x0(ne*l+j)*x0(ne*l+k);
                l=l+1;
            end
            k=k+1;
        end
        k=1;
        while k<=ne
            l=1;
            while l<=j-1
                |\Hilig|x0(ne*k+j)=x0(ne*k+j)-gsc(l)*x0(ne
                |\Hilig|*k+l);
                l=l+1;
            end
            k=k+1;
        end;
        zn(j)=0.0;
        k=1;
        while k<=ne
            zn(j)=zn(j)+x0(ne*k+j)^2;
            k=k+1;
        end
        zn(j)=sqrt(zn(j));
        k=1;
        while k<=ne
           |\Hiligh|x0(ne*k+j)=x0(ne*k+j)/zn(j);
           k=k+1;
        end
        j=j+1;
    end
%  update running vector magnitudes
    k=1;
    while k<=ne;
        c(k)=c(k)+log(zn(k));
        k=k+1;
    end;
%  normalize exponent
    k=1;
    while k<=ne
        LE(k)=c(k)/(t-t_start);
        k=k+1;
    end
%   print results
    if (mod(it,out)==0)
        fprintf('%10.2f %10.4f %10.4f %10.4f\n',[t,LE']);
    end;
    i=1;
    while i<=ne
        j=1;
        while j<=ne;
            x(ne*j+i)=x0(ne*i+j);
            j=j+1;
        end
        i=i+1;
    end;
    x=transpose(x);
    it=it+1;
%  display LEs
    for i=1:ne
        plot(t,LE(i),'.','Color',colors(i));
        %plot(t,LE(i),'.','markersize',1,'Color',colors(i));%for thiny draw
    end
end
xlabel('t','fontsize',16)
ylabel('LEs','fontsize',14)
set(gca,'fontsize',14)
box on;
line([0,t],[0,0],'color','k')
axis tight
\end{lstlisting}

Function $dark.m$

\begin{lstlisting}%[Caption={a},captionpos=b,escapechar=|,tabsize=1]

function f=dark(t,x)
%Output data must be a column vector
f=zeros(size(x));
%variables allocated to the variational equations
X= [x(4), x(7), x(10);
x(5), x(8), x(11);
x(6), x(9), x(12)];
%RF equations
f(1)=x(2)*x(3)-x(1);
f(2)=(x(3)-5)*x(1)-x(2);
f(3)=1-x(1)*x(2);
%Jacobian matrix
J=[-1,x(3),x(2);
x(3)-5,-1,x(1);
-x(2),-x(1),0];
%Righthand side of variational equations
f(4:12)=J*X;
\end{lstlisting}

\noindent\textbf{ERRATA}

Following the new (improved) variants of the codes at \cite{cod}, in \cite{dan1}

\noindent -instead (p.8):

\lstset{numbers=none,frame=none}
\begin{lstlisting}
run_FO_Lyapunov_p
\end{lstlisting}

\noindent it should be

\lstset{numbers=none,frame=none}
\begin{lstlisting}
run_LE_FO_p
\end{lstlisting}

\noindent -instead (p.8):

\lstset{numbers=none,frame=none}
\begin{lstlisting}
run_FO_Lypaunov_q
\end{lstlisting}

\noindent it should be:

\lstset{numbers=none,frame=none}
\begin{lstlisting}
run_FO_LE_q
\end{lstlisting}

\end{multicols}

\begin{figure}[ptbh]
 \includegraphics[width=.5\textwidth]{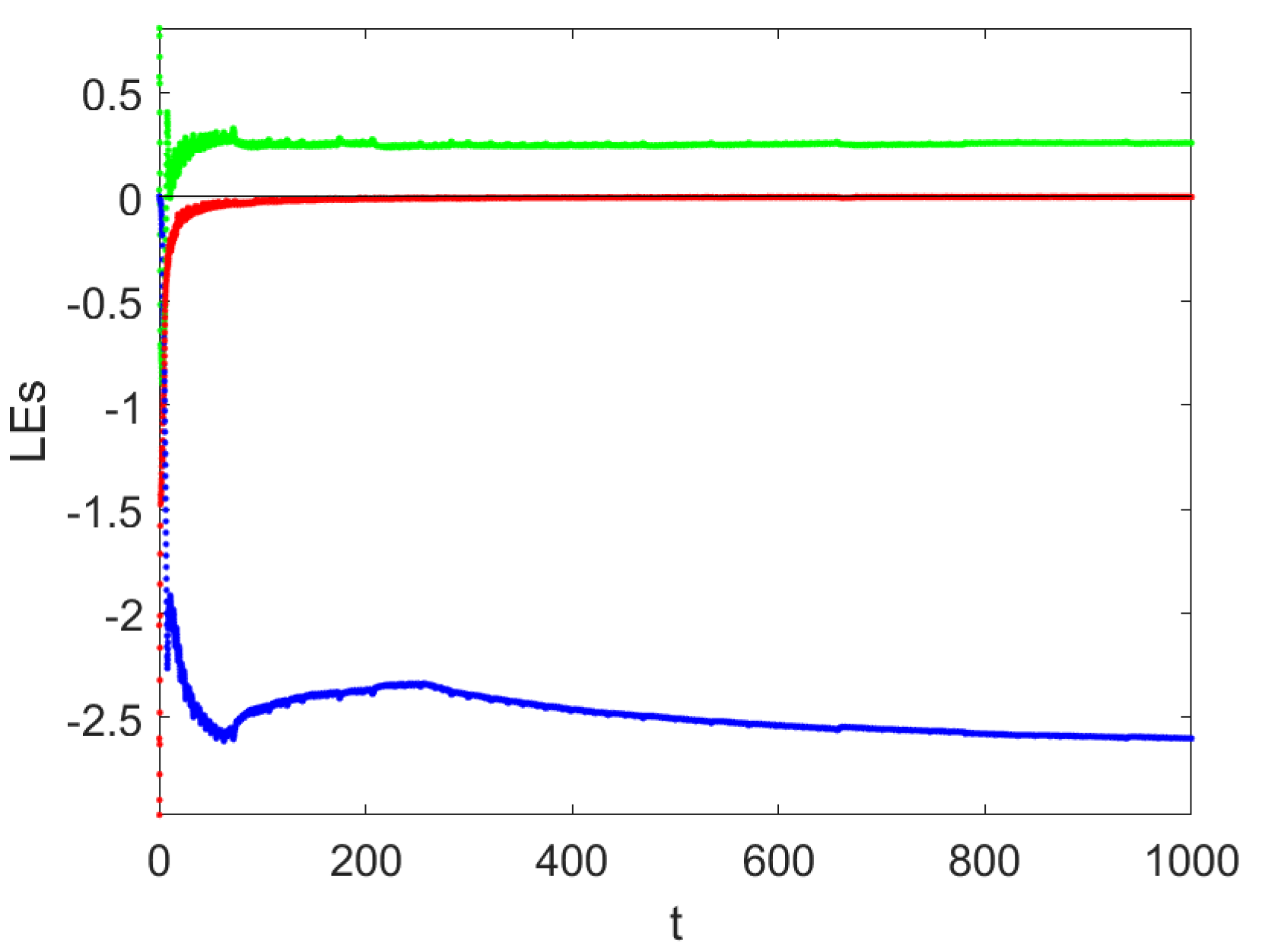}
\caption{Spectrum of LEs of the system \eqref{ecc}, obtained with $FO\_Lyapunov.m$ code.}
\label{figLE}       
\end{figure}

\begin{figure}[ptbh]
 \includegraphics[width=.5\textwidth]{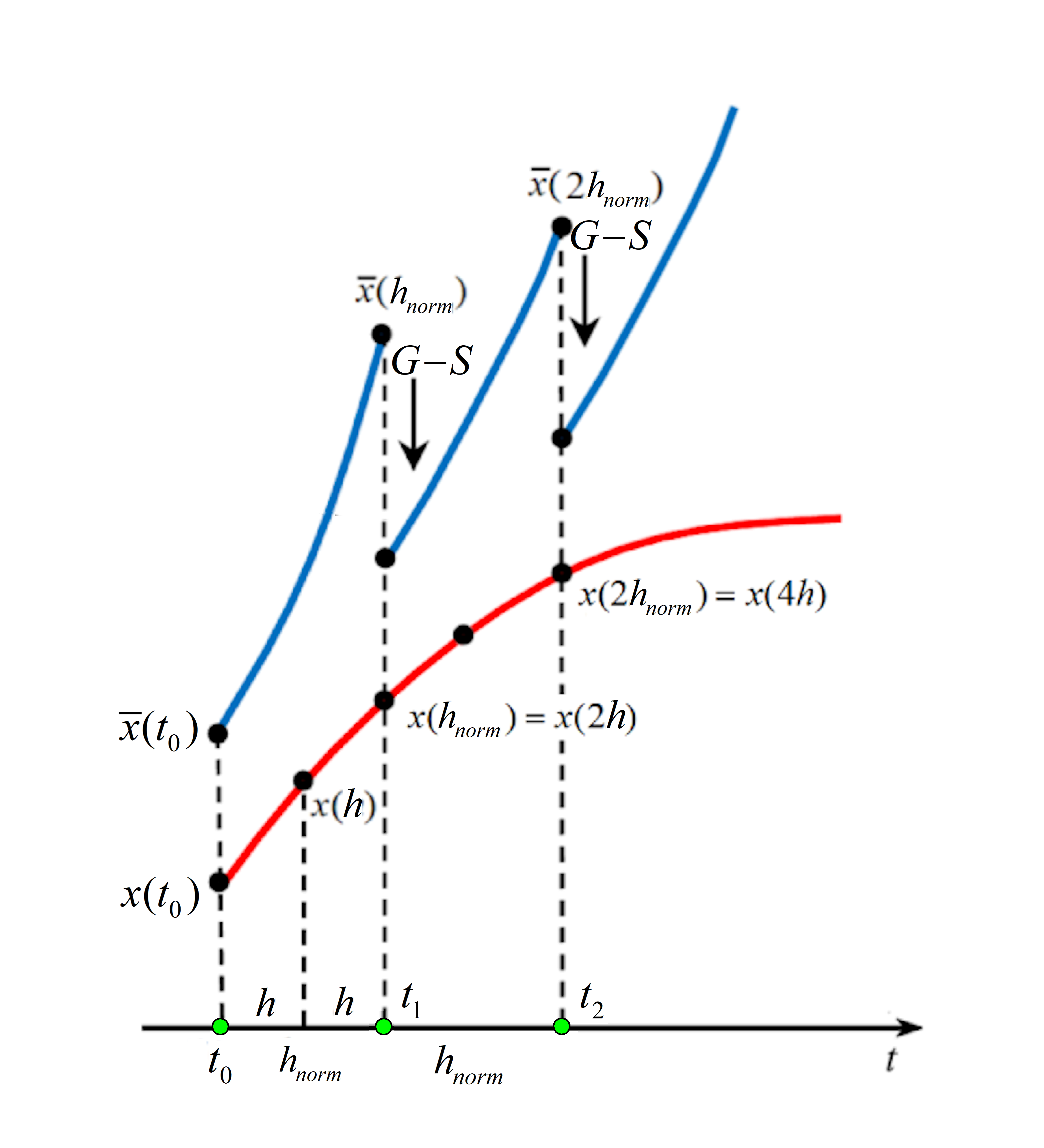}
\caption{Sketch revealing the impulsive action of G-S procedure. At every node $t_i, i=1,2,...$, when the numerical integration with fixed step-size $h$, acts on orthonormalization time intervals of length $t_{norm}$, is applied the G-S procedure.}
\label{fig5}       
\end{figure}
\begin{figure}
\noindent\begin{minipage}{.47\textwidth}
\renewcommand{\lstlistingname}{Algorithm}
\begin{lstlisting}[caption={Gram-Schmidt matlab code}, label={ddd},escapechar=|,tabsize=1]
function u=gramschmidt(v)
    [n,k] = size(v);
    u=zeros(n,k);
    u(:,1)=v(:,1)/norm(v(:,1));
    for i=2:k
        u(:,i)= v(:,i);
        for j = 1:i-1
          |\Hilight| u(:,i)= u(:,i)-(u(:,j)'*u(:,i))*u(:,j);
        end
        |\Hiligh|u(:,i)=u(:,i)/norm(u(:,i));
    end
end
\end{lstlisting}
\end{minipage}

   \end{figure}
\end{document}